\documentclass[11pt]{article}
\usepackage{hyperref}
	
	\newcommand{\blind}{0}
	
    \makeatletter
    \renewcommand\section{\@startsection {section}{1}{\z@}%
                                       {-3.5ex \@plus -1ex \@minus -.2ex}%
                                       {2.3ex \@plus.2ex}%
                                       {\normalfont\fontfamily{phv}\fontsize{16}{19}\bfseries}}
    \renewcommand\subsection{\@startsection{subsection}{2}{\z@}%
                                         {-3.25ex\@plus -1ex \@minus -.2ex}%
                                         {1.5ex \@plus .2ex}%
                                         {\normalfont\fontfamily{phv}\fontsize{14}{17}\bfseries}}
    \renewcommand\subsubsection{\@startsection{subsubsection}{3}{\z@}%
                                        {-3.25ex\@plus -1ex \@minus -.2ex}%
                                         {1.5ex \@plus .2ex}%
                                         {\normalfont\normalsize\fontfamily{phv}\fontsize{14}{17}\selectfont}}
    \makeatother
	\usepackage{amsmath}
	\usepackage{graphicx}
	\usepackage{subfig}
	\usepackage{enumerate}
	\usepackage{xcolor}
	\usepackage{natbib} %comment out if you do not have the package
	\usepackage{url} % not crucial - just used below for the URL
\begin{document}
		
			%%%%%%%%%%%%%%%%%%%%%%%%%%%%%%%%%%%%%%%%%%%%%%%%%%%%%%%%%%%%%%%%%%%%%%%%%%%%%%
		\def\spacingset#1{\renewcommand{\baselinestretch}%
			{#1}\small\normalsize} \spacingset{1}
		%%%%%%%%%%%%%%%%%%%%%%%%%%%%%%%%%%%%%%%%%%%%%%%%%%%%%%%%%%%%%%%%%%%%%%%%%%%%%%
		
		\if0\blind
		{
			\title{\bf Digitization of Raster Logs: A Deep Learning Approach}
			\author{M Quamer Nasim$^a$ $^b$, Narendra Patwardhan$^a$, Tannistha Maiti$^a$ and Tarry Singh$^a$ \\
			$^a$ Deepkapha.ai, Amsterdam, Netherlands \footnote{Corresponding author email: quamer.nasim@deepkapha.com}\\
             $^b$ Indian Institute of Technology, Kharagpur, India }
			\date{}
			\maketitle
		} \fi
		
		\if1\blind
		{

            \title{\bf \emph{IISE Transactions} \LaTeX \ Template}
			\author{Author information is purposely removed for double-blind review}
			
\bigskip
			\bigskip
			\bigskip
			\begin{center}
				{\LARGE\bf \emph{IISE Transactions} \LaTeX \ Template}
			\end{center}
			\medskip
		} \fi
		\bigskip
		
	\begin{abstract}Raster well-log images are digital representations of well-logs data generated over the years. Raster digital well logs represent bitmaps of the log image in a rectangular array of black (zeros) and white dots (ones) called pixels. Experts study the raster logs manually or with software applications that still require a tremendous amount of manual input. Besides the loss of thousands of person-hours, this process is erroneous and tedious. To digitize these raster logs, one must buy a costly digitizer that is not only manual and time-consuming but also a hidden technical debt since enterprises stand to lose more money in additional servicing and consulting charges. We propose a deep neural network architecture called VeerNet to semantically segment the raster images from the background grid and classify and digitize the well-log curves. Raster logs have a substantially greater resolution than images traditionally consumed by image segmentation pipelines. Since the input has a low signal-to-resolution ratio, we require rapid downsampling to alleviate unnecessary computation. We thus employ a modified UNet-inspired architecture that balances retaining key signals and reducing result dimensionality. We use attention augmented read-process-write architecture. This architecture efficiently classifies and digitizes the curves with an overall F1 score of 35\% and IoU of 30\%. When compared to the actual las values for Gamma-ray and derived value of Gamma-ray from VeerNet, a high Pearson coefficient score of 0.62 was achieved. 
	\end{abstract}
			
	\noindent%

	%\newpage
	\spacingset{1.5} % DON'T change the spacing!

\section{Introduction} \label{s:intro}
Well-logging is the process of taking measurements of various rock properties along the
length of the well down into the ground by drilling tools. The digital log responses are
functions of lithology, porosity, fluid content, and textural variation of formation. The well-logging parameters are used to derive lithofacies groups and facies-by-facies descriptions of rock properties. Before the advent of digital logging instruments, well-logging data were drawn on the parameter graph in curve format. Well-logging parameter graphs have many disadvantages: large size, ample memory space, and interference like gridlines. Therefore, it is necessary to convert well-logging parameter graphs into X-Y coordinates, where X represents parameter values and Y represents depth values. Raster logs are scanned copies of paper logs saved as image files. 
 
Well log data saved as depth-calibrated raster images provide an economic alternative to digital formats for preserving this valuable information into the future \citep{cisco1996raster}. Although often discarded after vectorization, raster imaged well logs may be the key to a global computer-readable format for legacy hardcopy data. This legacy data is stored on multiple media and contains information for various applications in addition to resource exploration and development, such as environmental protection, water management, global change studies, and primary and applied research.  

Experts such as geologists/reservoir engineers revisit and study these raster logs manually or with software applications requiring a tremendous amount of manual input. Besides the loss of thousands of person-hours, the existing process is erroneous and tedious. To digitize these raster logs and efficiently use them in conventional as well as unconventional analysis, one needs to buy an expensive digitizer which is a manual and time-consuming task but also there is a hidden technical debt since enterprises stand to lose more money in additional servicing and to consult charges. The commercially used logging curve digitization software, Neuralog, is based on SCTR (Scanning, Compressing, Tracing, and Rectifying). However, due to the interference of the background grid, this software frequently pauses during curve tracking.Several unsupervised computer vision methods have been implemented to digitize the log data embedded in the binary image.

Well-log digitization could be performed through two kinds of approaches: Pixel-based methods and non pixel-based methods. Pixel-based methods include the thinning process and the Global Curve Vectorization (GCV) method  \citep{zheng2005parallel,hilaire2001improving, nagasamy1990engineering}. The thinning method reduces the width of a line to only one pixel, leaving only the skeleton that can characterize its features. The main disadvantage of the thinning process is that it has a high time complexity, loses line width information, and is prone to deformation and wrong branches in the intersection area. GCV method is suitable for line processing but poor for point line processing  \citep{yuan2019digitization}.

Non-pixel-based methods mainly fall into two categories: contour-based and adjacency graph-based. The contour-based approach \citep{hori1993raster} is to extract the contour of the image first and then find the matched contour pairs. The adjacency graph method first applies run-length encoding to graphs, then analyses the segments and generates various adjacency graph structures, such as line adjacency graph (LAG) and block adjacency graph (BAG)  \citep{pavlidis2012algorithms}. \citep{li2003sctr} used SCTR (Scanning, Compressing, Tracing, and Rectifying) approach by employing the LAG data structure. \cite{yang2009pctr} improved the SCTR method and put 
forward the PCTR (Preprocessing, Compressing, Tracing, and Rectifying) method. \cite{yuan2019digitization} proposed an algorithm for erasing grid lines and reconstructing strokes in Chinese handwriting based on BAG. However, such methods are difficult to deal with the complex situation in well-logging parameter graphs, especially the analysis of nodes. \cite{yuan2019digitization} used morphological image processing and pixel statistics method to eliminate gridlines, isolating the curves and the gridlines. Then, the remaining grid lines and noise points are cleared according to the characteristics of the small size of their connected components. However, all these existing methods need manual intervention, which is not the desired option, especially when the paper logs have a huge size $>$ 10 MB. 
In the present study, we propose a novel transformer-based deep learning model named \emph{VeerNet} which employs self-attention mechanisms to identify individual curves from a single track. The straightforward design of transformers allows the processing of various modalities (e.g., image, video, text, and speech) using similar processing blocks. Transformers demonstrate high scalability to large-scale deep neural networks and large datasets. These strengths have led to exciting progress on several vision tasks using Transformer networks \cite{khan2021transformers}.

We train our model on synthetic data as well as on real raster logs. In the multi-segmentation model of digitizing a track with three curves, VeerNet performs with a precision of 0.94, 0.48 and 0.39 respectively. Whereas the VeerNet trained on real data has a precision of ~0.6 for three curves. 

\section{Work Flow in commercial software vs our proposed solution} \label{s:sec1}
\begin{figure}[h]
\includegraphics[width=12cm]{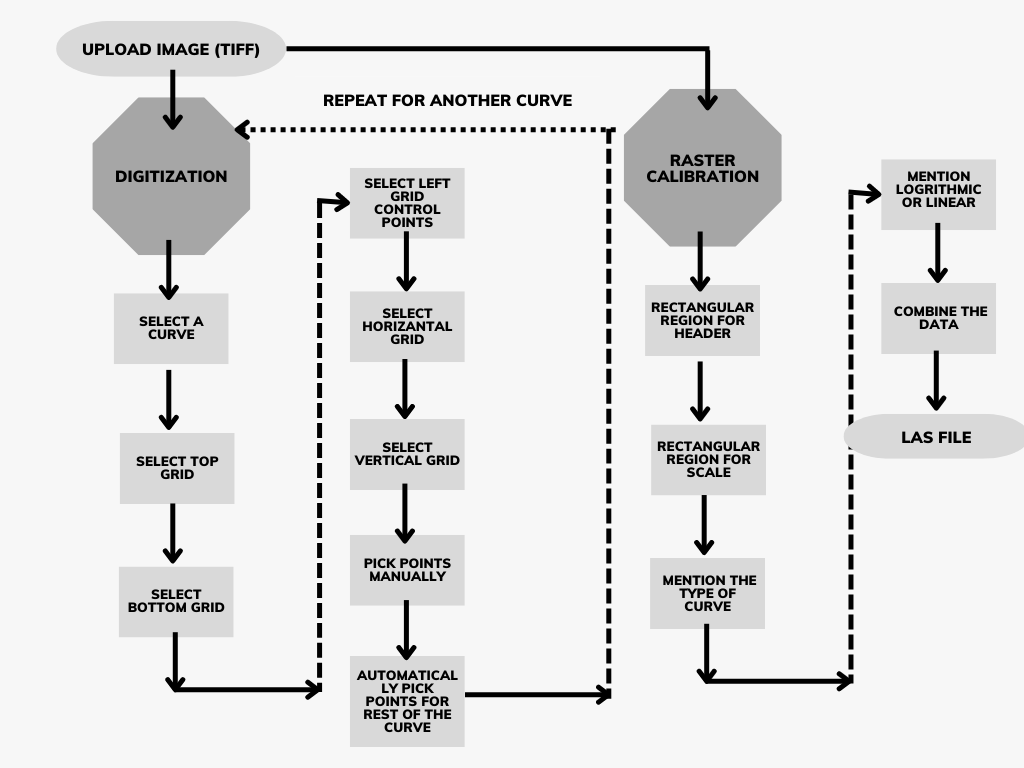}
\centering
\caption{Overview of digitization steps performed in Neuralog Software. In the software user needs to provide depth points, left and right scale points, depth grid for each cure separately. }
\label{neuralog}
\end{figure}

\begin{figure}[h]
\includegraphics[width=12cm]{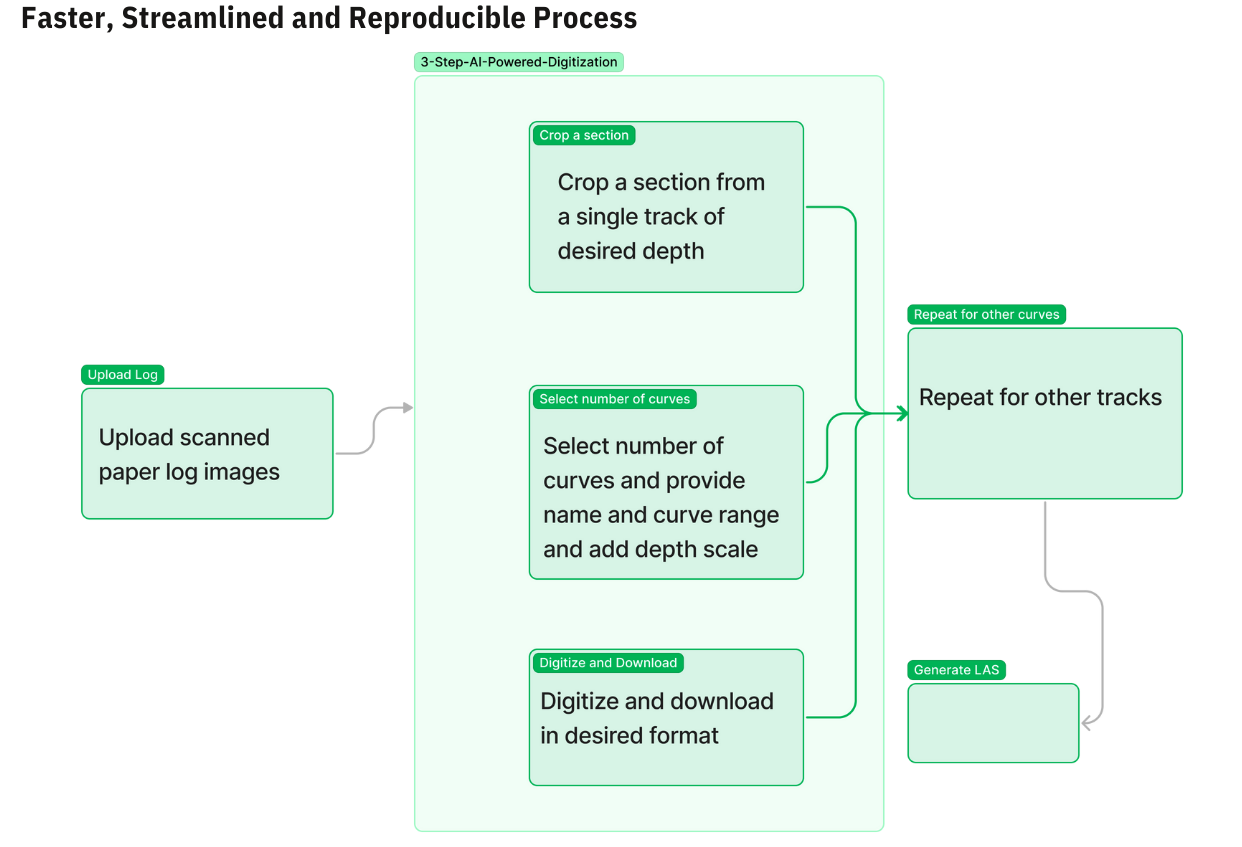}
\centering
\caption{Overview of digitization steps performed in our solution. The three-step begins with (a) uploading a scanned paper and cropping a selected section of the log image (b) Providing the number of curves present in the cropped section and the scale of the selected curves and depth scale (c) Generating CSV/las.}
\label{digitizer}
\end{figure}

Fig.\ref{neuralog} illustrates the overall workflow of the existing system for digitizing well-log graphs. This workflow is from the recognized logging curve digitization software - 
\href{http://www.neuralog.com}{Neuralog}
Due to interference with the background grid, this software frequently pauses during curve tracking. The software interface is based on a ribbon structure and includes a few modules. The steps include (a) raster calibration and (b) digitization. During the raster calibration, the user must provide a rectangular region that captures the header and scale areas. Also, a set of points is defined to capture the depth of the log tracking. User input is also required to determine the left and right axis values and the type of scale, whether logarithmic or linear. For the digitization ribbon, one needs to identify the track’s top and bottom, define the image track with width, and add depth points and grids. Finally, the user must pick points in the curve, and the software’s auto-trace functionality will trace
the rest of the curve. Our proposed solution (Fig.~\ref{digitizer}) operates on minimal manual intervention. The user is not required to provide left or right input points; they don’t need to specify the width of the track. \textit{VeerNet} architecture does not have any metadata requirements related to depth grids. Therefore, the algorithm can efficiently differentiate between grids and curves. The user only needs to provide a cropped raster or paper log section.

\begin{figure}[hbt!]
\includegraphics[width=12cm]{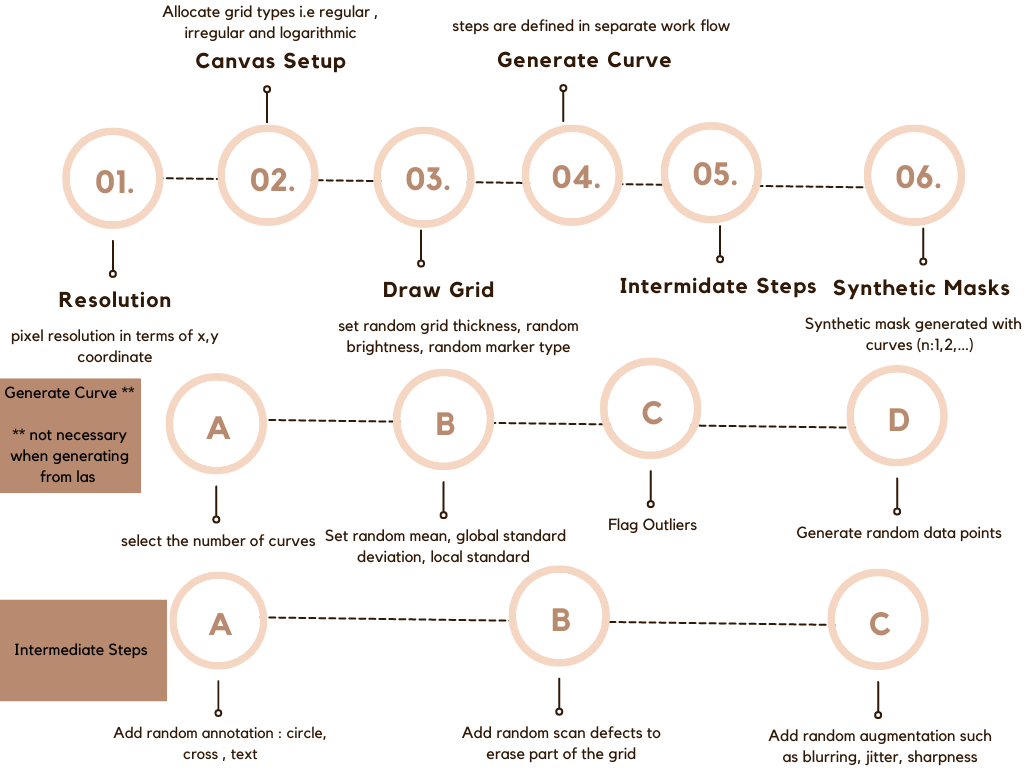}
\centering
\caption{Pipeline for synthetic dataset generation. It begins with setting the canvas resolution, drawing the grid on the canvas, selecting the number of curves that should be present in the synthetic well-log image and set the statistics of the curve and generating the random data points. Finally, plot the generated data points on canvas.}
\label{syntgen}
\end{figure}
\section{Dataset: Synthetic and Real log curves } \label{s:sec2}

\subsection{
\emph{Synthetic Dataset}}
\label{s:methods.1}
Reconstruction of well log curves through machine learning techniques like the random forest, SVM \citep{akinnikawe2018synthetic}, and deep neural network-based prediction models \cite{kim2020generation} is used to generate synthetic logs. These synthetic curves are generated for target fields and require domain dependence. Here we aim to develop a generalized model without any specific targeted oil field. We hypothesized that the curves in a particular track would have random overlapping and wrapping factors. The synthetic curves are generated based on mean and standard deviation, adding random noise. Our dataset consists of two well-log curves per image. The synthetic curves were developed based on parameters such as mean, global standard deviation, local standard deviation, etc. The pipeline consists mainly of 4 steps namely a) Canvas preparation b) Curve generation c) Noise and annotation addition and d) Masks \& Log generation. The pipeline generates a single track and only includes 3 tracks per image.

In canvas preparation, we first set the resolution of this digital canvas and then add grids to it (Fig~\ref{syntgen}). Grids can be linear with a regular scale and a logarithmic grid with a log scale. To closely match raster images, the thickness and brightness of grid lines are randomly assigned. This step mimics the erosion introduced in pdf or image prints. Masks are generated similarly without gridlines. 

The curve generation phase begins by defining the number of curves generated for a particular track (1, 2, 3, ..., n). In real generally have 1) Local variation 2) Global variation. We use random mean, global standard deviation, and local std dev to mimic real well-log curves. Next, a parameter defines the number of global variations (default =10). This parameter sets local variation in each segment and 10 global variations in the entire curve. Finally, a flag is defined to introduce sudden changes in the value of the curve, a behavior similar to real well-logs. 

In the third step, we introduce several defects and annotations synthetically, usually observed in real logs because of scan defects. Augmentation, such as random blurring, random sharpness, and color jitter, were added to increase the noise in the dataset. In addition, we included random annotations such as circles, crosses, random text to mimic geologist's markings on the logs. The final step is adding all the curves generated in steps 2 and step 3 to the empty canvas. We generated 15014 images with 2 curves and 20000 images with 3 curves respectively.

\subsection{
\emph{Real Dataset}} \label{s:methods.2}
The real dataset was generated from las and raster image files obtained from \href{https://www.rrc.state.tx.us/}{Texas RRC}. The real dataset was generated using about 1000 las files. The well-log curves used in the real dataset are - gamma ray, caliper, SP, shallow resistivity, medium resistivity, deep resistivity, neutron porosity and density logs. First, we generated mini-batches from the extracted well-log curves of las files. Then, a Gaussian process regression fit was applied to the mini-batches, which generated about 100 distributions for the curves. Next, the distribution was randomly selected and sampled. Finally, we created the real dataset with data sampled from the distribution. To treat the NaN values in the las files, we implemented two methods (a) fill the NaN value with a constant number and (b) Remove the NaN values. A detailed analysis of the real dataset is available in the Table~\ref{tab:real data}.
\begin{table}[hbt!]
\centering
\begin{tabular}{|p{1.5cm}|p{5cm}|p{2cm}|p{2cm}|}
\hline
Track & Well-log Curve & Number of curves in each track & Number of Images Generated \\
\hline
Track 1 & SP, Gamma Ray, Caliper & 1 & 1600 \\
\cline{3-4} &  & 2 & 1600 \\
\cline{3-4} & & 3 & 1600 \\
\hline
Track 2 & Shallow Resistivity, Medium Resistivity and Deep Resistivity & 1 & 1408 \\
\cline{3-4} & & 2 & 1280 \\
\cline{3-4} & & 3 & 1280 \\
\hline
Track 3 & Density log and Neutron Porosity Log & 1 & 1280 \\
\cline{3-4} & & 2 & 1280 \\
\hline
\end{tabular}
\caption{\label{tab:real data} Distribution of images based on two/three tracks and the respective well-log curves used in model training and validation. The dataset is generated from processed real data.}
\end{table}

\section{Architecture}
\label{s:numerical}
\begin{figure}[!ht]
\includegraphics[width=12cm]{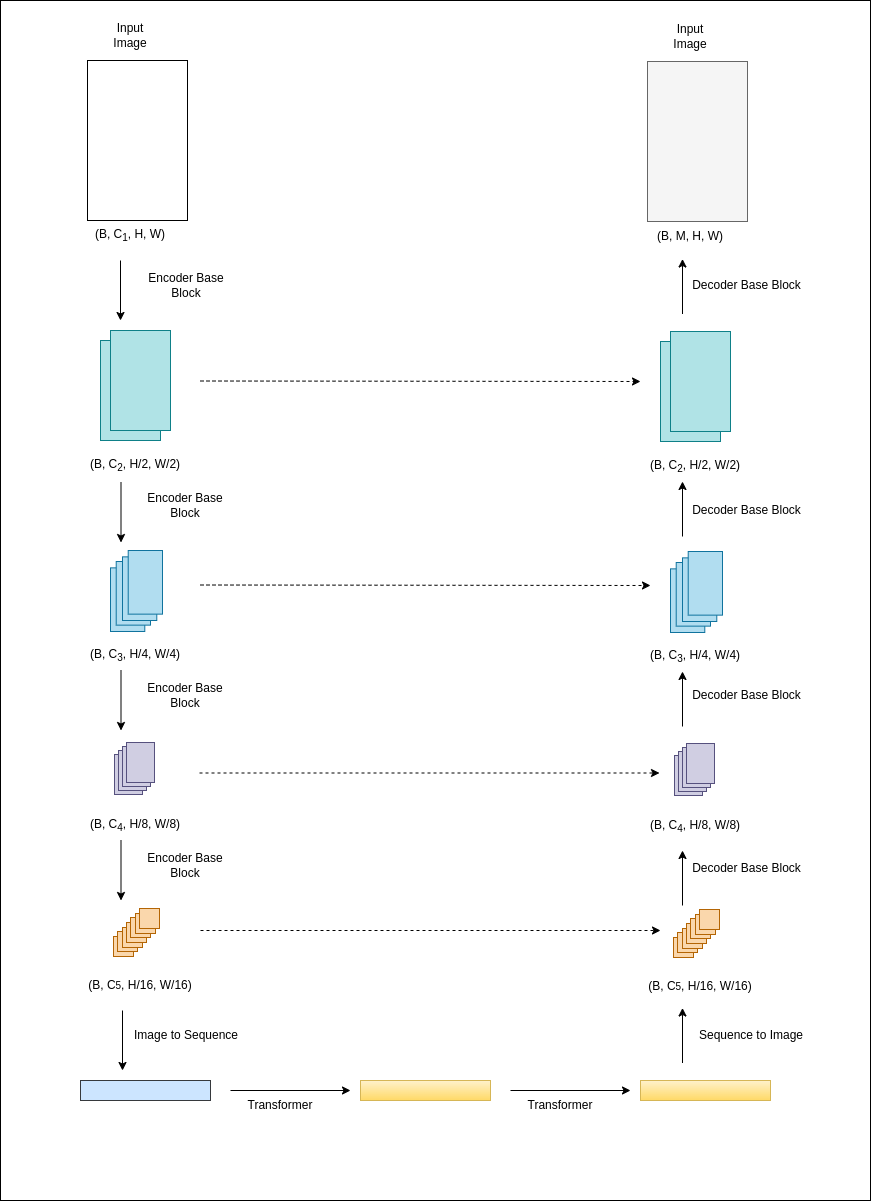}
\centering
\caption{Transformer augmented U-Net }
\label{taunet}
\end{figure}

Raster logs have a substantially greater resolution as compared to images traditionally consumed by image segmentation pipelines, which restricts the usage of transfer learning due to memory requirements. As the input has a low signal-to-resolution ratio, we require rapid downsampling to alleviate unnecessary computation. We thus employ a modified UNet-inspired architecture called \textit{VeerNet} that strikes a balance between retaining key signal and reducing result dimensionality.
VeerNet does this using attention augmented read-process-write architecture: the input is downsampled in the encoder using a series of 4 blocks that consist of a 2D convolution that reduces the spatial resolution by a factor of 2, group normalization, and GELU nonlinearity. It is the tokenized and fed transformer blocks \cite{vaswani2017attention} that process and refine the internal representation learned by the encoder. In the end, the sequence is reformed to match corresponding encoder level resolutions with a decoder. The decoder blocks follow the sequence of bilinear upsampling, 2D convolution, group normalization, and RELU activation. Residual connections improve the signal strength at each decoder level (shown in Fig~\ref{taunet}). The final decoder block produces spatial masks that signify the presence of corresponding log curves.
This approach inherits the best of both UNet and modern transformer-based models: It keeps inference time and memory requirements low by exploiting inductive biases for efficient downsampling with convolution and allows learning rich, global context-aware representations by use of transformers.

\section{Training}\label{s:training}
We trained on (a) synthetic and (b) real datasets comprising 35000 and 10000 images, respectively. The images consist of a single track that can either have 2 or 3 well-log curves. We performed several experiments with various loss functions and also changed the number of transformers (i.e. 4, 5, and 6) and a maximum number of epochs of 250 to train the models. The learning rate of the experiments varies from 1.5e-3 to 2.5e-3. We used a cosine learning rate for the learning rate scheduler for all the runs. We trained our models on 5 NVIDIA A100-SXM-80GB GPUs for an average of 14 hours.

The VeerNet model comprises an Encoder-Decoder architecture. The Encoder consists of Residual blocks, which support information flow from shallower to deeper layers. Five residual blocks encode the image in a feature map $1/32th$ the size of the original image. Then, followed by Encoder, we’ve two transformer layers, essentially made up of an attention layer. The attention weights are computed for the input feature map and produce an output vector with encoded information on how each pixel should attend to all other pixels in the image. Finally, followed by the Transformer layer, we’ve five decoder layers, each comprising an upscaling operation and a convolution operation to attain the exact output size as given in the input. The architecture will always produce the same size mask as the original image. We then use post-processing steps on the predicted masks to generate CSV files which is the digitized value of the well-log curve or curves. 

\subsection{Loss Functions}
The various loss functions used in hyper-parameter tuning are described here:
\subsubsection{Dice Loss}
Dice Loss optimizes networks based on the dice overlap coefficient between the predicted segmentation result and the ground truth annotation. Thus can effectively alleviate the imbalance between the foreground and background \cite{zhao2020rethinking}. 

\subsubsection{Tversky Loss}
Tversky Loss is based on the Tversky similarity index \cite{salehi2017tversky}. Adjusting the hyperparameters of this index allow placing emphasis on false negatives in training a network that generalizes and performs well in highly imbalanced data as it leads to high sensitivity. 
\subsubsection{Lovasz Loss}
Lovasz loss is a loss function for multiclass semantic segmentation \cite{berman2018lovasz} . The Lovasz loss achieves direct optimization of the mean intersection-over-union loss. It improves the IoU.
\subsubsection{Focal Loss}
Focal loss is just an extension of the cross-entropy loss function that would down-weight easy examples and focus training on hard negatives. So to achieve this, researchers have proposed: $(1- p_t)\gamma$ to the cross-entropy loss, with a tunable focusing parameter $\gamma\geq0$.
\subsubsection{Sparse Cross Entropy Loss}
Sparse categorical cross entropy is used when the classes are mutually
exclusive (e.g. when each sample belongs precisely to one
class) \cite{totakura2020prediction}. It works on integers, which is true, but they must be class indices, not actual values. This loss computes logarithm only for output index, which ground truth indicates. So, when the model output is, for example, [0.1, 0.3, 0.7] and the ground truth is 3 (if indexed from 1), the loss computes the only logarithm of 0.7. It calculates the logarithm once per instance and omits the summation, which leads to better performance.

\subsection{Evaluation Metrics}
\subsubsection{Pearson coefficient ($r_p$)}
$r_p$ or bivariate correlation measures linear correlation between two variables X and Y for finite sample sizes \cite{de2016comparing}. It measures the strength of association between two variables and the direction of their relationship. In terms of the strength of relationship, the value of the correlation
coefficient varies between +1 and -1. A value of ±1indicates a perfect degree of association between the two variables. As the correlation coefficient value goes towards 0, the relationship between the two variables will be weaker. The direction of the relationship is indicated by the sign of the coefficient; a + sign
indicates a positive relationship and a – sign indicates a negative relationship.
\subsubsection{$p-value$}
p-value to estimate the linear relationship
between two variables. In this study, a $p-value <$0.05
refers to a statistically significant difference between variables and supports that two samples did not come from the same distribution. A $p$ indicates no statistically significant difference, and two samples come from the same distribution.

\section{Results}\label{s:conclusion}

We treat this problem in two stages (a) multi-class classification problem to identify the
well-log curves from the background and separate them into individual curves (b) regression
analysis of individual curves to find the goodness of fit. 80\% of the total instance 10000 were kept for training, while 20\% of the synthetic and real dataset instances were used for validation based on the
results reported. 
The model takes the user’s paper log as an input and applies filters to extract feature maps. Further, the model tries to reduce the paper log’s feature map size. The middle layer gives attention to signals in the paper log, ignoring other aspects such as grid lines and annotations. This derived information is used to expand the log feature map into its original shapes so that everything else but the signal is removed in the image. So in the end we have an image which has only signals present in them without any grid lines or other noises. Once we have this signal image or mask, we extract a 1D signal from the same and save it in CSV/LAS files. 

The study’s central hypothesis is based on a low signal/noise ratio. While efforts were made to introduce noise similar to that observed in raster paper logs, the technique still has limitations. VeerNet achieves an overall F1 score of $>$35\% when applied to the real dataset and an overall F1 score of $>$ 30\% for the synthetic dataset (refer to Table~\ref{tab:real} and Table~\ref{tab:synthetic}).The maximum IoU score for synthetic and
real ranges between $>$26\% and $>$30\%, respectively. In the first stage of the evaluation of mask, the performance metrics used were Intersection over Union (IoU) and F1 Score. IoU is a metric that evaluates Deep Learning algorithms by estimating how well a predicted mask
matches the ground truth data. F1-score sums up the predictive performance of a model by combining two otherwise competing metrics— precision and recall. The experiments were performed with five different loss functions. Lovasz loss performed best amongst all the loss
functions.

\begin{table}[hbt!]
\centering
\begin{tabular}{ |p{4cm}|p{2cm}|p{2cm}| }
 \hline
 Loss Function and Transformer Numbers & F1  & IoU \\
 \hline
 Lovaz, 5   & \textbf{0.35}    &\textbf{0.30}  \\
 \hline
 \emph{Sparse Cross Entropy, 3} (4,6) &   \textbf{0.34} (0.33,0.33)  & \textbf{0.30} (0.30,0.29) \\
 \hline
 Focal, 6  & 0.32 & 0.28 \\
 \hline
 Tversky, 6 (4)    &0.24 (0.22) & 0.20 (0.22)\\
 \hline
 Dice, 4 (5)   &0.25 (0.23) & 0.22 (0.19)\\
 \hline
\end{tabular}
\caption{\label{tab:real} Results of VeerNet on real dataset with model variation based on loss and transformer numbers. The best model is reported for Lovaz and Sparse Cross Entropy loss functions based on F1 score. The numbers in the braces indicate the results obtained by training for the different number of transformers.}
\end{table}
For the synthetic dataset, experiments were performed with varying learning rates and learning loss ranging from 0.06 to 0.008. The best F1 score for Focal loss was reported with $lr$ of \emph{0.05} and the least F1 score for \emph{0.01} with \textit{lr-warmup-decay} of \emph{0.001}. Other experiments were performed using default learning rates for respective loss functions. Lovasz Loss and Focal loss displayed the best results in terms of IoU (26\%, 23\%) and F1 (30\%, 28\%) scores.

In the experiments for the real dataset, experiments were performed with a default learning rate (refer to Table~\ref{tab:real}). We used the number of transformers as a hyper parameter to study the effect on F1 scores. No significant metrics change was observed (Table 5). The study on the impact of the number of transformers is inconclusive. SCE
(sparse cross entropy) and Dice 3 transformers give better F1 than 5, 4, and 6, respectively. Whereas, for Tversky, using six transformers provide a better score. Lovasz Loss and SCE loss displayed the best results Table~\ref{tab:real}).

\begin{table}[hbt!]
\centering
\begin{tabular}{ |p{4cm}|p{3cm}|p{3cm}| }
 \hline
 Loss Function  & F1  & IoU \\
 \hline
 Lovasz   & 0.30   &0.26  \\
 \hline
 Sparse Cross Entropy &   0.26 & 0.24  \\
 \hline
 Focal   & 0.28  & 0.23  \\
 \hline
 Tversky    &0.21 & 0.15 \\
 \hline
 Dice   &0.19 & 0.13\\
 \hline
\end{tabular}
\caption{\label{tab:synthetic} Results of VeerNet on synthetic dataset with model variation based on loss. The best model is reported for Lovaz and Focal loss functions based on F1 score.}
\end{table}
\begin{table}
\centering
\begin{tabular}{ |p{6.5cm}|p{2cm}|p{2cm}| }
 \hline
 Loss Function and Curve Specification & Pearson Coefficient & $p-value$ \\
 \hline
 Lovasz and GR  & 0.62    &0.0   \\
 Sparse Cross Entropy and GR &   0.27  & 0.0  \\
 Lovaz and CALI &-0.21 & 0.0 \\
 Sparse Cross Entropy and CALI    &0.12 & 0.04\\
 \hline
\end{tabular}
\caption{\label{tab:real-curve} Statistical variants for VeerNet. Pearson coefficient and $p-value$ determine how well VeerNet performed when compared to real data.}
\end{table}

 \begin{figure}[!ht]
\centering
    \subfloat[]{\includegraphics[width=0.24\linewidth]{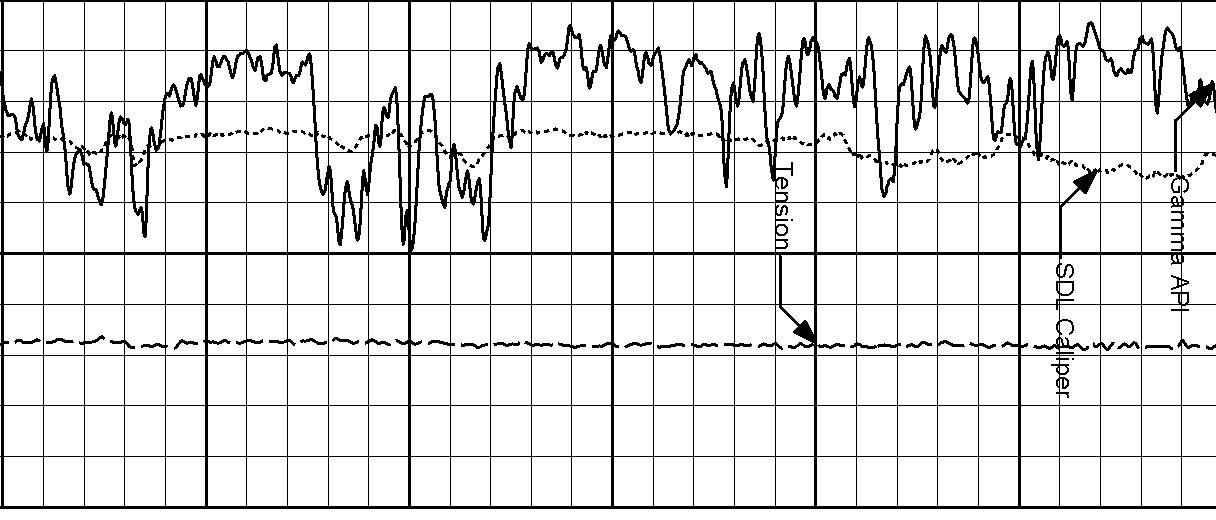}}
\hfil
    \subfloat[]{\includegraphics[width=0.3\linewidth]{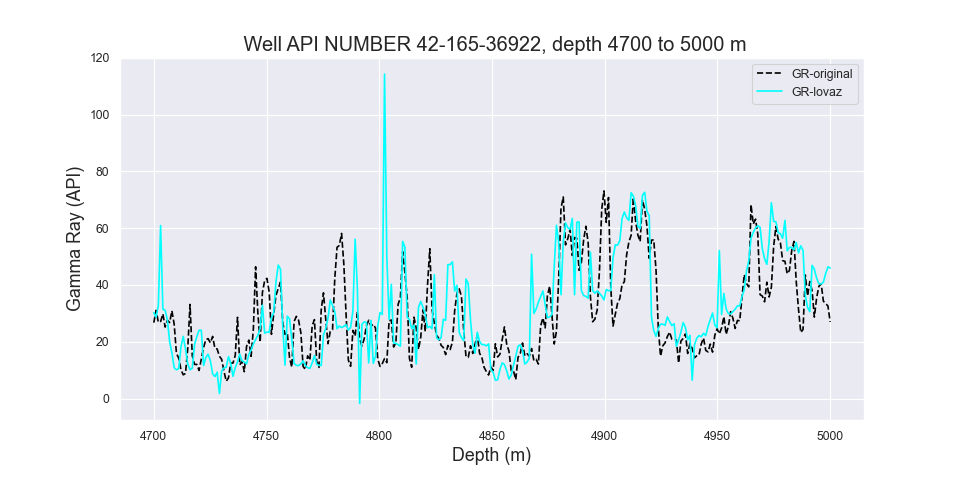}}
    \subfloat[]{\includegraphics[width=0.3\linewidth]{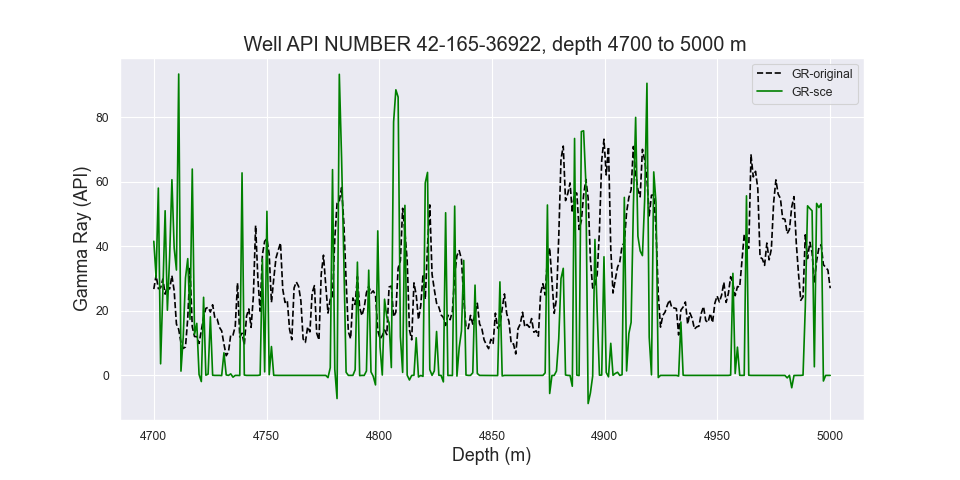}}

    \subfloat[]{\includegraphics[width=0.25\linewidth]{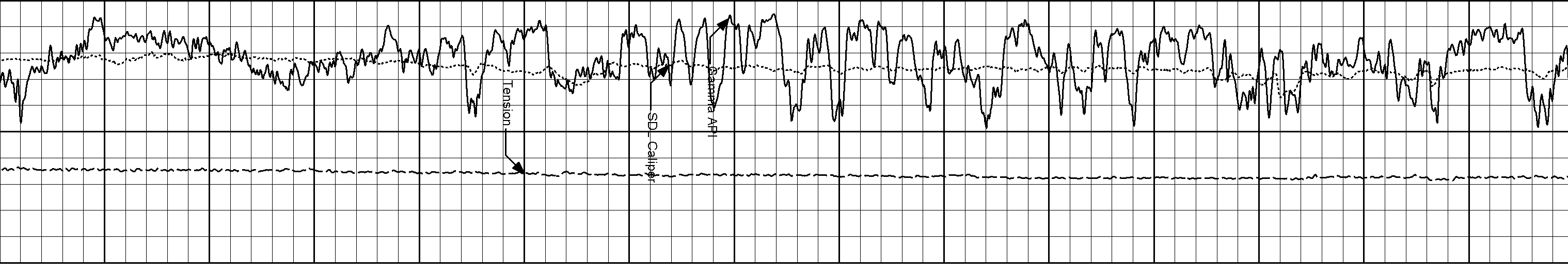}}
\hfil
    \subfloat[]{\includegraphics[width=0.3\linewidth]{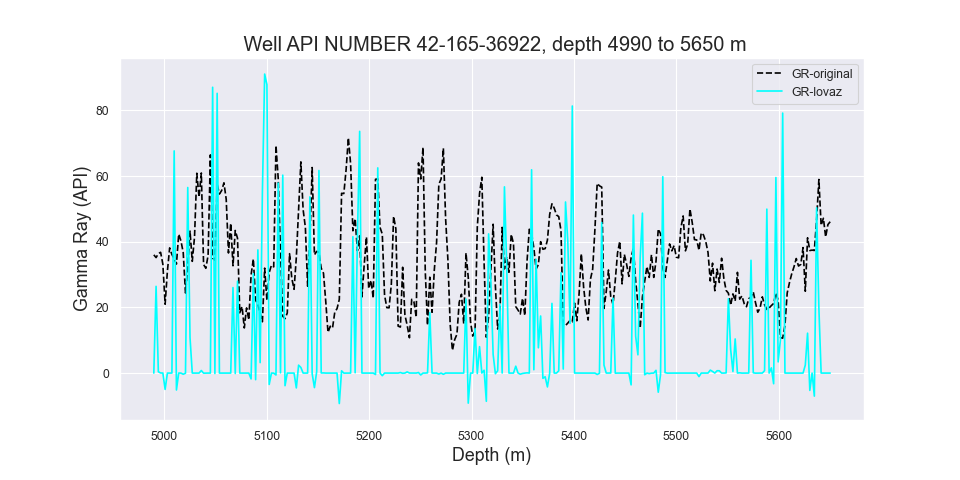}}
    \subfloat[]{\includegraphics[width=0.3\linewidth]{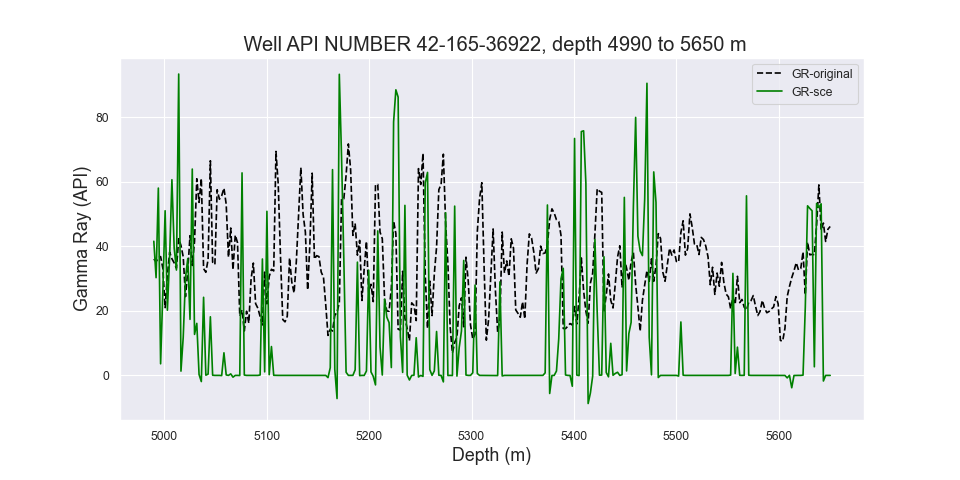}}

    \subfloat[]{\includegraphics[width=0.25\linewidth]{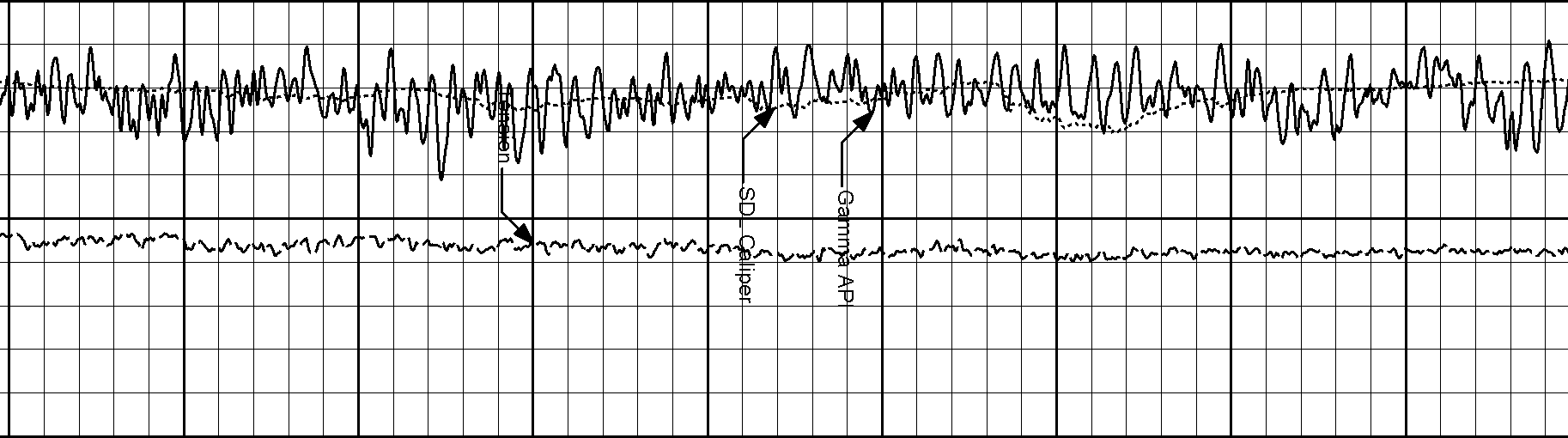}}
\hfil
    \subfloat[]{\includegraphics[width=0.3\linewidth]{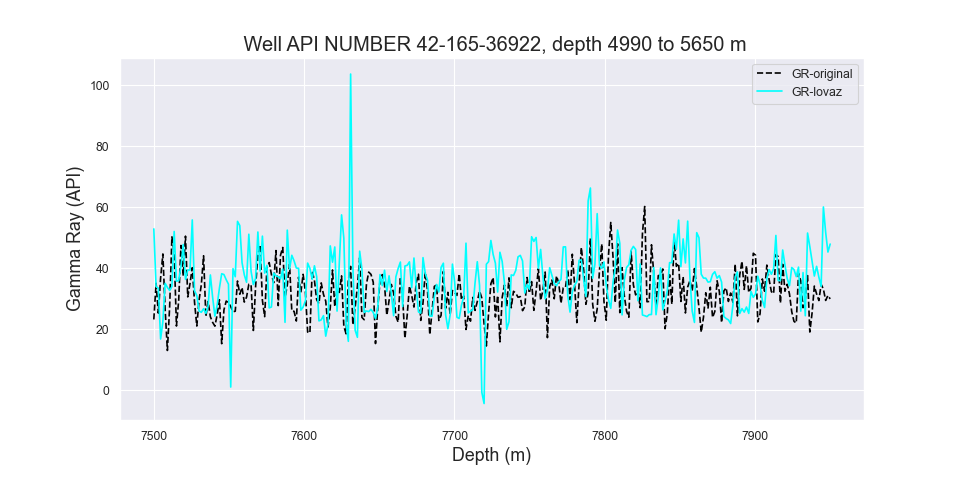}}
    \subfloat[]{\includegraphics[width=0.3\linewidth]{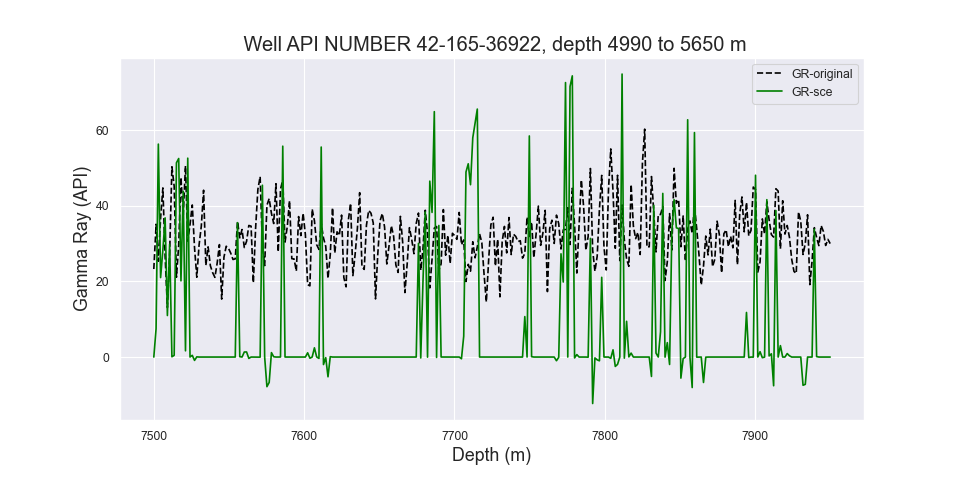}}
\caption{Comparison between las value of well with API number: \emph{42-165-369222} and and well name: \emph{UNIVERSITY 6-13 No. 1}. (a): Section from raster tif image at depth 4700 to 5000 m (d): Section from raster tif image at depth 4990 to 5650 m
(g): Section from raster tif image at depth 7500 to 7950 m. (b), (e) and (h): represents the fit between las data and gamma-ray value processed from VeerNet model with Lovaz as loss function (c), (f) and (i): represents the fit between las data and gamma-ray value processed from VeerNet model with sparse cross entropy as loss function}
    \label{fig:gamma-ray}
    \end{figure}

\begin{figure}[!ht]
\centering
    \subfloat[]{\includegraphics[width=0.24\linewidth]{4700-5000.png}}
\hfil
    \subfloat[]{\includegraphics[width=0.3\linewidth]{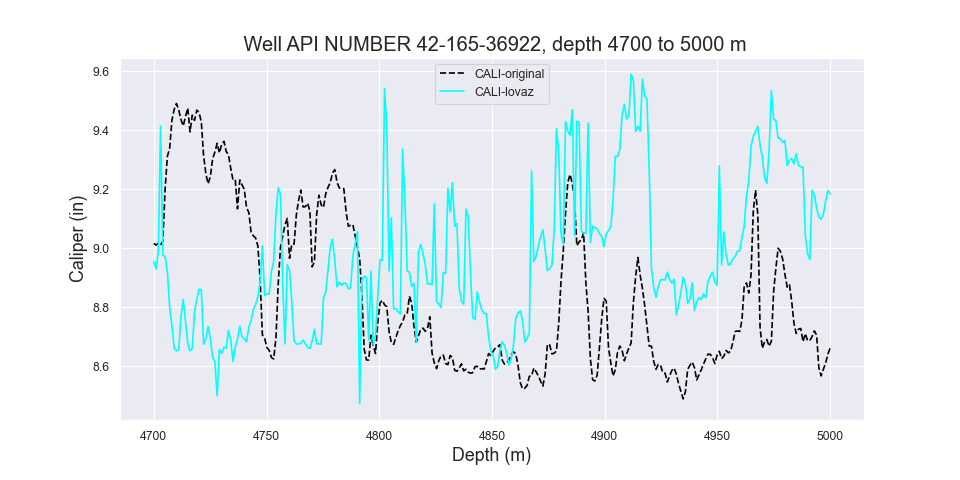}}
    \subfloat[]{\includegraphics[width=0.3\linewidth]{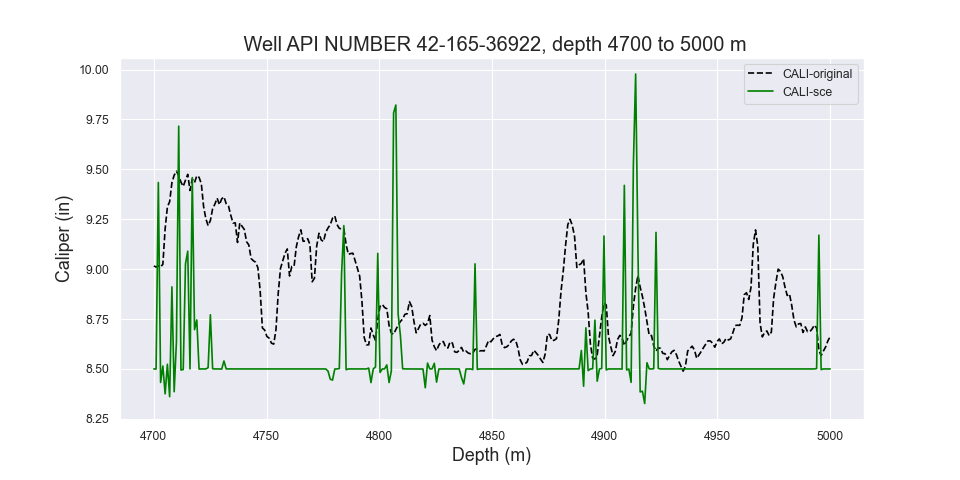}}

    \subfloat[]{\includegraphics[width=0.25\linewidth]{4990-5650.png}}
\hfil
    \subfloat[]{\includegraphics[width=0.3\linewidth]{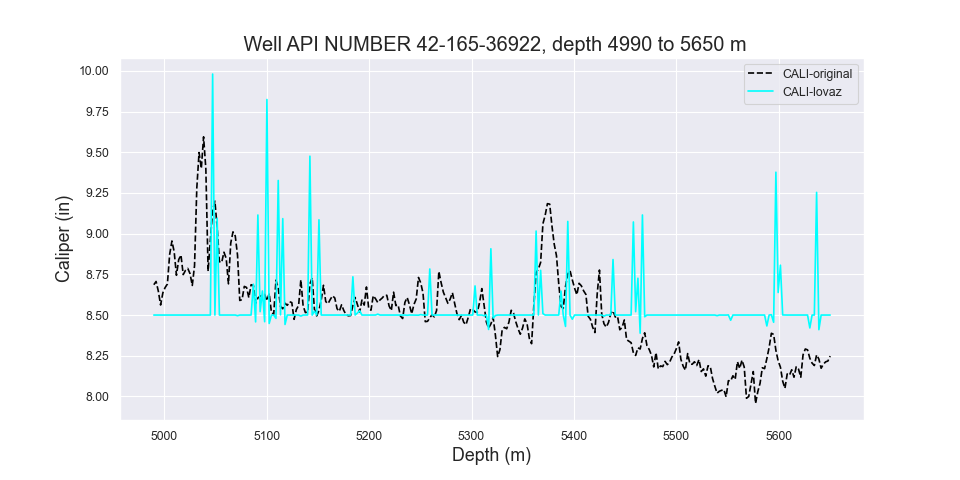}}
    \subfloat[]{\includegraphics[width=0.3\linewidth]{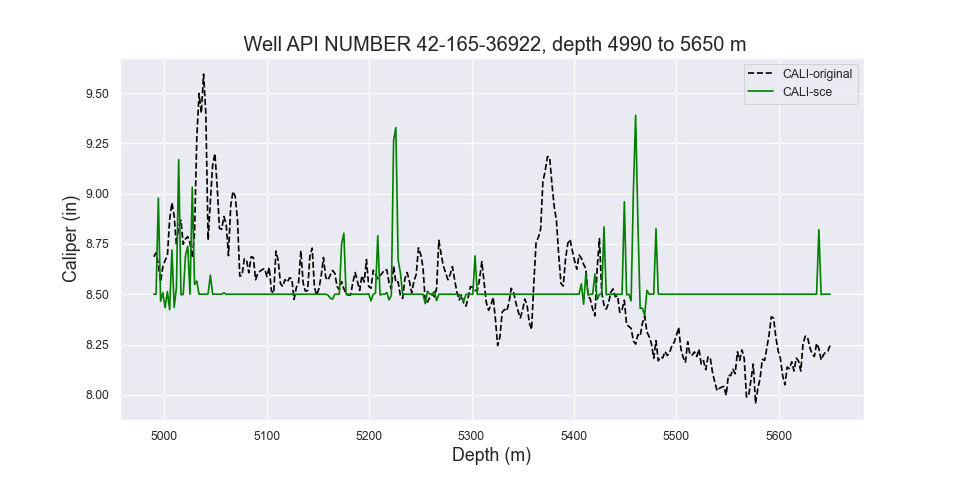}}

    \subfloat[]{\includegraphics[width=0.25\linewidth]{7500-7950.png}}
\hfil
    \subfloat[]{\includegraphics[width=0.3\linewidth]{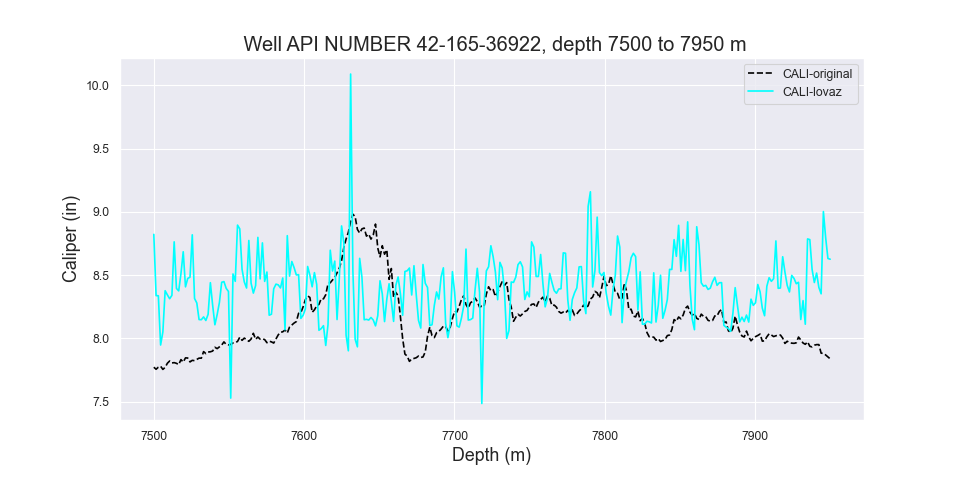}}
    \subfloat[]{\includegraphics[width=0.3\linewidth]{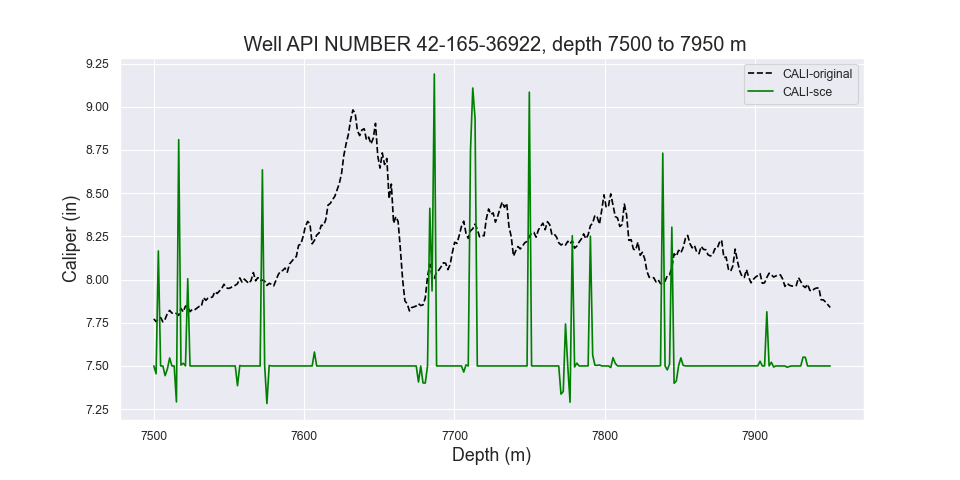}}
\caption{Comparison between las value of well with API number: \emph{42-165-369222} and and well name: \emph{UNIVERSITY 6-13 No. 1}. (a): Section from raster tif image at depth 4700 to 5000 m (d): Section from raster tif image at depth 4990 to 5650 m
(g): Section from raster tif image at depth 7500 to 7950 m. (b), (e) and (h): represents the fit between las data and caliper value processed from VeerNet model with Lovaz as loss function (c), (f) (i): represents the fit between las data and caliper value processed from VeerNet model with sparse cross entropy as loss function.}
    \label{fig:caliper}
    \end{figure}
Experimental results obtained from VeerNet models on Gamma Ray and Caliper are listed in Fig~\ref{fig:gamma-ray} and Fig~\ref{fig:caliper}, which demonstrate that the proposed model fits the actual well log curve with certain goodness of fit. To experimentally explore the potential fit of the digitized data with las data we implemented two model variants with loss as Lovasz and Sparse Cross Entropy.  
In statistical studies, the null hypothesis is a default hypothesis that a quantity (typically the difference between two situations) to be measured is zero (null). In this scenario, the null hypothesis is to determine if there is an indication that the samples from calculated and original are derived from different distributions. The p-value is the probability of obtaining test results at least as extreme as the results observed, assuming that the null hypothesis is correct. A $p-value$ $<$ 0.05 is sufficient to reject the null hypothesis and conclude that a significant difference between the two distributions does exist. To measure the similarity derived gamma ray from Lovasz loss and SCE loss we calculate  Pearson coefficient $r_p$ and $p-value$ value for the model with Lovasz loss and SCE for Gamma-ray (Table~\ref{tab:real-curve}) are high, indicating the derived value and the original value las can correlate. Also $p-value$ $<$ 0.05, hence the distributions are statistically significant therefore rejecting the null hypothesis. 
To measure the similarity derived for caliper log from Lovasz loss and SCE loss we calculate  Pearson coefficient $r_p$ and $p-value$. The $r_p$ value for the model with Lovasz loss has a negative correlation while for SCE a low positive correlation of 0.12 (Table~\ref{tab:real-curve}), indicating the derived value and the original value from las remains inconclusive. Also the $p-value$ $<$ 0.05 is observed for both model variants with Lovasz and SCE,hence the distributions are statistically significant therefore rejecting the null hypothesis.

\section*{Conclusion}
In conclusion, our proposed solution for digitization of raster well-log images is simple and
less manual intervention than the existing techniques. The solution is fast and scalable
too. We have introduced a deep learning model, VeerNet that can efficiently classify well-
log curves and can achieve a classification accuracy $>$35\%. The model can perfectly
identify well-log curves from the background grid, improving existing technology. We demonstrate digitized gamma-ray values from various sections of the well compared to actual las, which generates a high Pearson coefficient of 0.62. This value indicates the overall improved goodness of fit to real data. Since this is a new study, there is no baseline model to compare the results. We also introduced new techniques to generate synthetic well-log data using Gaussian process regression, where the generated synthetic data is basin-independent.

One limitation of using one track is that the user needs to provide a cropped section from the front end. This functionality could include an error in manual analysis in providing the scale value. To overcome these difficulties, we propose (1) Keeping all the 3 or 4 tracks of the raster well log images and (2) designing OCR architecture to automatically read scales of the well-log curves rather than providing them manually. In the next iteration of VeerNet we will provide a detailed analysis of all the hyperparameters discussed above.

\bibliographystyle{chicago}
\spacingset{1}
\bibliography{IISE-Trans}
	
\end{document}